\documentclass[12pt, letterpaper, twoside]{article}

\usepackage{authblk}
\usepackage{graphicx}
\usepackage{dcolumn}
\usepackage{bm}
\usepackage{amsmath}
\usepackage{amssymb}
\usepackage{siunitx}
\usepackage[version=4]{mhchem}
\usepackage{ulem}
\usepackage{upgreek}
\usepackage{gensymb}
\usepackage{tabularx}
\newcolumntype{b}{X}
\newcolumntype{s}{>{\hsize=.27\hsize}X}
\usepackage{xcolor}
\definecolor{mypink}{RGB}{255, 0, 181}
\definecolor{myblue}{RGB}{0, 149, 255}
\definecolor{mycyan}{RGB}{0, 255, 249}
\definecolor{mydarkgreen}{RGB}{0, 128, 128}
\definecolor{mygreen}{RGB}{0, 128, 0}

\begin{document}
\title{Sugar-Pucker Force-Induced Transition in Single-Stranded~DNA}

\author[1,2,*]{X. Viader-Godoy}
\author[1]{M. Manosas}
\author[1]{F.Ritort}
\affil[1]{%
 Departament de F\'{i}sica de la Mat\`eria Condensada, Universitat de Barcelona, Diagonal 647, 08028 Barcelona, Spain
}%
\affil[2]{%
 Department of Physics and Astronomy, University of Padova, via Marzolo 8, 35131 Padova, Italy
}%

\affil[*]{\textit{ xavier.viadergodoy@unipd.it}}

\date{\today}
\maketitle         

\begin{abstract}
The accurate knowledge of the elastic properties of single-stranded DNA (ssDNA) is key to characterize the thermodynamics of molecular reactions that are studied by force spectroscopy methods where DNA is mechanically unfolded. Examples range from DNA hybridization, DNA ligand binding, DNA unwinding by helicases, etc. To date, ssDNA elasticity has been studied with different methods in molecules of varying sequence and contour length. A dispersion of results has been reported and the value of the persistence length has been found to be larger for shorter ssDNA molecules. We carried out pulling experiments with optical tweezers to characterize the elastic response of ssDNA over three orders of magnitude in length (60--14 k bases). By fitting the force-extension curves (FECs) to the Worm-Like Chain model we confirmed the above trend:the persistence length nearly doubles for the shortest molecule (60 b) with respect to the longest one (14 kb). We demonstrate that the observed trend is due to the different force regimes fitted for long and short molecules, which translates into two distinct elastic regimes at low and high forces. We interpret this behavior in terms of a force-induced sugar pucker conformational transition (C3$’$-endo to C2$’$-endo) upon pulling ssDNA.
\end{abstract}

\section{Introduction}
\label{S:Intro}
DNA is a nucleic acid polymer encoding the genetic information of organisms.  In its most common conformation DNA forms a double-stranded helix (dsDNA) stabilized by hydrogen bonds and base-stacking interactions~\cite{Florian1999,Sponer2001,Kamenetskii2006}. However DNA can also be present in single-stranded form (ssDNA) where the strands remain dissociated. ssDNA occurs in many biological processes requiring the reading of the genetic information, such as DNA replication~\cite{Alberts2003}, transcription~\cite{Griffiths2000} and repair~\cite{Friedberg2005}. The double-stranded and single-stranded polymer forms present different biochemical and mechanical properties: whereas dsDNA is pretty stable and stiff~\cite{Finzi1992}, ssDNA is more reactive and flexible~\cite{Smith1996}.  
Force-spectroscopy techniques have been widely used to study biochemical and enzymatic processes involving DNA, such as unzipping of DNA structures~\cite{Huguet2010PNAS,Huguet2017NAR}, binding of ligands to DNA~\cite{Vladescu2007,Manosas2017}, unwinding of DNA catalyzed by helicases~\cite{Mehta1999,Johnson2007,Manosas2013}, etc. In these experiments extracting useful information (e.g. free energies of DNA conformations, position of ligand binding along DNA or determining the translocation rate of DNA motors) requires a detailed knowledge of the mechanical response of ssDNA. Many studies have addressed the elastic properties of dsDNA~\cite{Baumann1997,seol2007dsDNA,Shon2019}, however, much less is known about the elasticity of ssDNA. Single-molecule force-spectroscopy techniques have allowed the measurement of their elastic response fitting it to the elastic Worm-Like Chain (WLC) and the Freely-Jointed Chain (FJC) models. The value of the persistence length of ssDNA, i.e., the distance over which the polymer is deformed by thermal forces,  is short (in the nanometer range) being comparable to the interphosphate crystallographic distance ($\sim$6 \AA). This fact makes the elastic response of ssDNA equally adjustable with a fit of the extensible FJC (with the Kuhn length and the stretching modulus as fitting parameters, and imposing the interphosphate crystallographic distance) or the inextensible WLC (with persistence length and contour length per base as fitting parameters). In comparison, dsDNA has a much larger persistence length ($\sim$50 nm), 150 times larger than the interphosphate distance \mbox{($\sim$3 \AA)}. The disparity of the two lengthscales underlines the importance of bending stiffness in dsDNA and the enthalpy gain upon straightening DNA by pulling. Consequently dsDNA cannot be fit well by the extensible FJC over a wide range of forces and the WLC must be used instead~\cite{Finzi1992,marko1995stretching}.

In contrast to the case of dsDNA, for ssDNA a large dispersion in the values of the WLC elastic parameters (persistence length $p$ and contour length per base $l$) has been reported from different experimental techniques and sequences~\cite{Smith1996,ClausenSchaumann2000,murphy2004probing,adamcik2006observation,Doose2007,Manohar2008,bosco2014elastic}. In particular it has been shown that base stacking affects the elasticity of ssDNA and ssRNA with different purine/pyrimidine content~\cite{PhysRevE.70.020902,PhysRevLett.99.018302,mcintosh2014}. Moreover, single molecule force studies have reported a systematic dependence of $p$ and $l$ of ssDNA on the molecular length, with larger $p$ and shorter $l$ values in short molecules (a few tens of bases)~\cite{murphy2004probing,Woodside2006PNAS,Chen2012,alemany2014determination} as compared to longer ones (a few hundreds or thousands of bases)~\cite{Smith1996,bosco2014elastic,Croquette2001,McIntosh2011,DeLorenzo2015,Saleh2009PRL,ClausenSchaumann2000}. In comparison, this finite length trend is the opposite of that observed in dsDNA where $p$ decreases for shorter molecules~\cite{seol2007dsDNA,forns2012handles} .
Furthermore, such a dependence of $p$ with molecular length in ssDNA is quite unexpected as $p$ spans at most two bases along the phosphate backbone, which is much lower than the polymer's length. How then the value of $p$ can be influenced by such a large scale?

\begin{figure}[h!]
\includegraphics[width=0.98\linewidth]{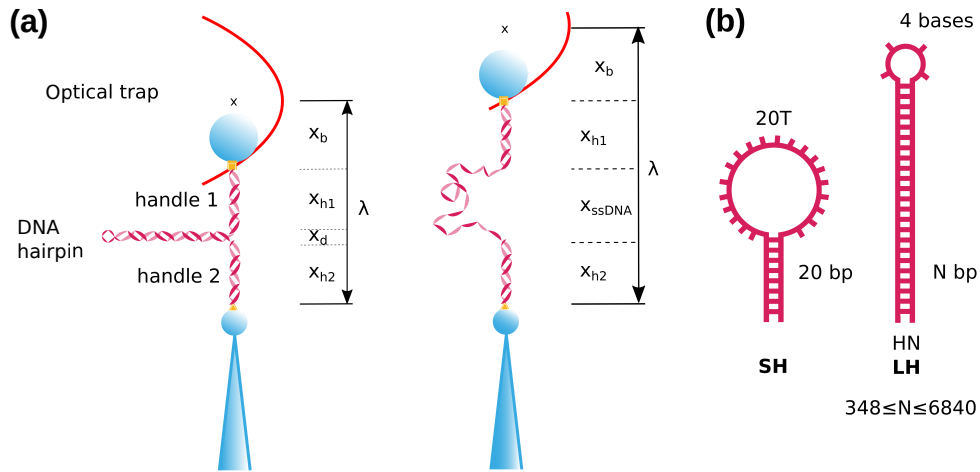}
\caption{\scriptsize{Experimental and molecular setup. ({\bf a}) A folded DNA hairpin is inserted between two dsDNA handles. The molecular construct is attached between two beads, one is held by air suction on the tip of a micropipette and the other is optically trapped by two counter-propagating lasers (schematically depicted as a (vertically oriented) harmonic potential). Upon stretching the molecular construct, the hairpin unfolds at a given force (\textbf{right}). $\lambda$ is the distance of the optical trap relative to the bead in the pipette. ({\bf b}) Schematic depiction of the two types of hairpins used in this work. The short hairpin, SH (20 bp), consists of a stem of 20 basepairs, with a loop of 20 bases \mbox{(1 guanine + 19 thymines).} The long hairpins ($\geq$348 bp), LHs, have a stem of $N$-bp ending in a~tetraloop.}}
\label{fig:1}
\end{figure}

Here, we investigate the elastic response of ssDNA molecules ranging from tens to tens of thousands of bases, over two-three orders of magnitude in total molecular length $L$. $L$ is proportional to the contour length per base and the number of bases of the molecule, $n$, $L=n\times l$. 
We pull the molecules with optical tweezers and measure the force-extension curves (FECs). These can be fit to the inextensible-WLC model, either at high forces (>15 pN) for the long molecules and at low forces (<15 pN) for short molecules. We find that the values of the elastic parameters $p,l$ of the inextensible-WLC depend on the force regime, rather than on molecular length. This demonstrates that the molecular length dependence of $p$ and $l$ reported in previous works is just an artifact of the different force regimes over which the elastic response was fitted. Interestingly, an extensible-WLC (that includes an stretching modulus) fits the data throughout the range of forces.~We interpret the extensibility of the WLC as a force-induced non-cooperative transition between two sugar pucker conformations~\cite{saengerprinciples2013,sinden1994dna} of the ssDNA bases. The North conformation (C3$'$-endo, A form) of the sugar pucker is favoured at low forces, whereas the South conformation (C2$'$-endo, B form) is favoured at high forces. Each conformation has specific parameters $p,l$ in the inextensible-WLC model. In particular, the contour length per base equals to the crystallographic interphospate distance of the corresponding sugar pucker conformation.

\section{Results}

We have carried out pulling experiments with laser optical tweezers (Materials and Methods) on DNA hairpins, generating ssDNA by mechanically unfolding them (Figure~\ref{fig:1}a). Two types of hairpin (several long and one short) are used (Figure~\ref{fig:1}b): Long hairpins (LH), denoted as HN, have a stem of N basepairs ($348 \leq {\rm N}\leq 6840$) and a tetraloop; the short hairpin (SH) sequence contains a 20 bp stem and a 20-bases loop. For manipulating the hairpins using optical tweezers hairpins a molecular construct containing the hairpin and molecular spacers (handles) has been prepared. Details of the synthesis can be found in Materials and Methods and in the Supplementary Material.

\begin{figure}[ht!]
\centering\includegraphics[width=0.8\linewidth]{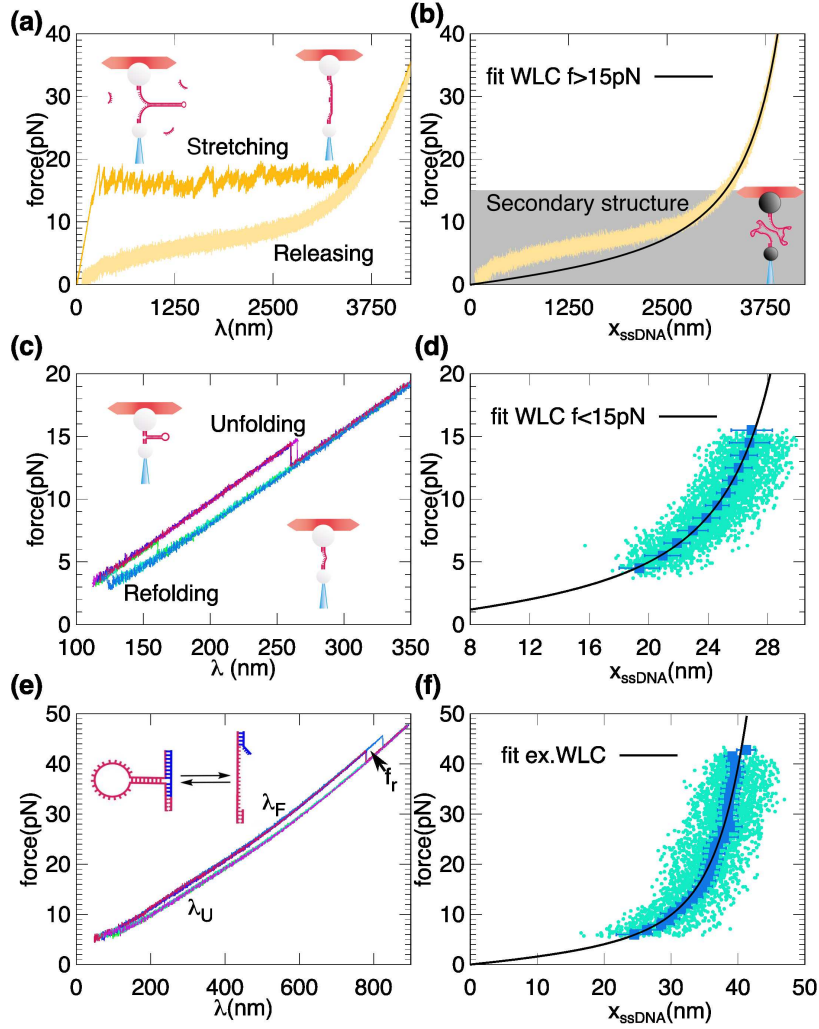}
\caption{\scriptsize{FDCs and FECs for long and short DNA hairpins. ({\bf a}) Stretching (upper sawtooth pattern) and releasing  (lower smooth pattern) FDC for a LH (H3594). ssDNA is obtained during the releasing process from long molecules by using the BLO method described in~\cite{bosco2014elastic}. ({\bf b}) Representative FEC for H3594. The gray area indicates the force range at which the secondary structure is formed and hence the fit using Equation~(\ref{eq::wlc_marko_siggia}) (black curve) is done only for high forces ($f>15$pN). ({\bf c}) Unfolding (upper branch of the FDC) and refolding (lower branch of the FDC) for the SH. ({\bf d}) Representative points of a FEC obtained for the SH. The blue squares are the averaged FEC obtained from 100 unfolding and refolding trajectories (cyan dots), using Equation~(\ref{eq:ssDNA_short}). The black curve is the fit {of} the inextensible WLC model (Equation~(\ref{eq::wlc_marko_siggia})) {to the averaged data}. ({\bf e}) Same as in ({\bf c}) but using the BSO method, with an overhang complementary to the second handle of 15 bases (blue oligo in the schematic depiction). ({\bf f}) Same results as in ({\bf d}) but obtained using the BSO method (Equation~(\ref{eq:ssDNA_short_block})). FEC data cover a larger force range as compared to ({\bf d}). The black curve is a fit {of} the extensible WLC model {of (}Equation~(\ref{eq::wlc_extensible})) {to the averaged data}.} \label{fig:2}}
\end{figure}

Hairpins were mechanically and repeatedly unfolded and refolded by moving the optical trap between minimum and maximum force values. Figure~\ref{fig:2}a shows force-distance curves (FDCs, where force is plotted versus the trap position $\lambda$) for LHs. FDCs show the characteristic unzipping sawtooth pattern consisting of a sequence of force rips and slopes. Above $\sim$15 pN, the hairpin is fully unzipped and the FDC shows the elastic response of the ssDNA (Figure~\ref{fig:2}a). We used the blocking loop oligonucleotide (BLO) method (Materials and Methods). In this method a short oligonucleotide that is complementary to the loop region is flowed into the chamber during the unzipping experiment to prevent the rezipping of the native hairpin. In order to measure the ssDNA elasticity we extracted the force-extension curve (FEC, where the force is plotted as a function of the molecular extension $x_{\rm ssDNA}$), as described in Materials and Methods. The FEC shows a shoulder characteristic of secondary structure formation~\cite{Croquette2001,bosco2014elastic} that deviates from the ideal elastic response \mbox{(Figure~\ref{fig:2}b,} grey region).

\begin{figure}[h!]
\includegraphics[width=1.0\linewidth]{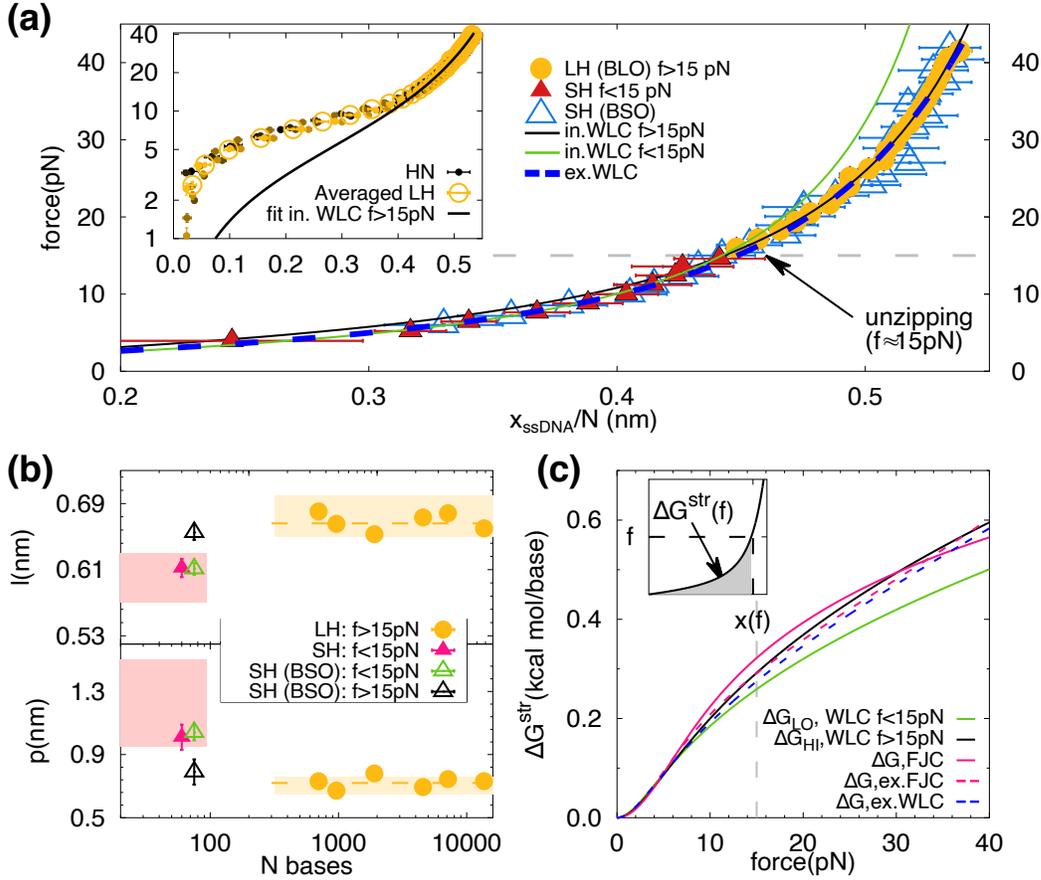}
\caption{\scriptsize{ssDNA elasticity ({\bf a}) {Averaged} FECs over LHs data (yellow circles), SH (red triangles), SH using the BSO method (empty blue triangles). Extensions, $x_{\rm ssDNA}$, are re-scaled by the total number of nucleotides, $N$. Green and black curves are the fits of the FEC to Equation~(\ref{eq::wlc_marko_siggia}) below and above 15 pN, respectively. The horizontal dashed line represents the division between the two force regimes. The blue dashed line is a fit to the extensible WLC {(}Equation~(\ref{eq::wlc_extensible}){)} throughout the whole force range. (Inset) Filled circles represent the re-scaled FECs of all the long hairpins (HN, $348\leq N\leq6840$, from darker to lighter color). The empty circles represent the averaged curve of all the studied lengths $N>348$ bp. ({\bf b}) Elastic parameters (contour length per base, $l$, and persistence length, $p$) extracted from LHs (yellow circles) and SH (red triangles). The yellow horizontal dashed lines are the average {values obtained} for all LHs. The empty triangles are the fitting parameters extracted from the fits shown in panel (a): green for $f< 15$ pN, black for $f>15$ pN. The magenta and yellow bands correspond to the range of elastic parameters reported in~\cite{camunas2016reviewnucleic} for short (tens of bases) and long (thousands of bases) ssDNA. ({\bf c}) Stretching energy per base as a function of the force, obtained from fitting the SH (BSO) data to different models. The magenta curves correspond to the values obtained from the FJC model (continuous line, $c=0.570(4)$ nm, $b=1.36(5)$ nm) and the extensible FJC (dashed line, $c=0.512(10)$ nm, $b=1.75(10)$ nm, $S=360(70)$ pN). The dashed grey vertical line at 15 pN indicates the typical unzipping force. (Inset) Schematic depiction of the the free energy of stretching, shown as the grey area below the black curve describing the ssDNA elasticity.}}
\label{fig:3}
\end{figure}
Therefore the ideal ssDNA elastic response for LHs can only be measured above a certain force ($f\sim$ 10--15 pN). On the other hand, the {SH FDCs show} force rips at high forces ($\sim$15 pN) and at low forces ($\sim$4 pN) during the stretching and releasing parts of the pulling cycle, respectively (Figure~\ref{fig:2}c). From the {FDCs}  we extract the ssDNA molecular extension at a given force from the difference in trap position between the folded and unfolded force-distance branches (Figure~\ref{fig:2}d, see Materials and Methods). {The main advantage of using a hairpin with a large loop of 20 b is that it decreases the refolding force, allowing for a wider force range at which the folded and unfolded branches coexist.} As we can see in Figure~\ref{fig:2}d, measurements of the ssDNA {elasticity} are limited to the force range set by the average unfolding and refolding forces (4--15 pN). To expand the range of measured forces, we implemented the blocking splint oligonucleotide (BSO) method that increases the force range in which the hairpin is folded (Figures~\ref{fig:2}e,f, see Materials and Methods). In this method a short oligonucleotide hybridizes to the hairpin fork region, acting as staple increasing kinetic stability of the hairpin.

In Figure~\ref{fig:3}a we show the {force} plotted versus the molecular extension normalized by the number of bases {for the different molecules studied}. Data includes results for the SH (Figure~\ref{fig:1}b, left) and for the LHs (Figure~\ref{fig:1}b, right) and cover different range of forces, $f<15$ pN (SH red triangles) and $f>15$ pN (LHs yellow circles), respectively. The blue triangles are the data obtained with the BSO method for the SH and cover the whole range of forces.

The LHs {FECs} deviate from the ideal elastic behaviour below 15 pN due to secondary structure formation (Figures~\ref{fig:2}b and~\ref{fig:3}a, inset). However, LHs data collapse into a single curve throughout the force range (Figure~\ref{fig:3}a, inset). Therefore, data for all LHs above 15 pN have been averaged and shown as yellow circles in the main figure. Notably, all FECs of ssDNA elasticity collapse into a single curve (Figure~\ref{fig:3}a, main), demonstrating the absence of molecular length effects in the explored range of lengths (60--14 kb).   
We fitted the data shown in Figure~\ref{fig:3}a to the inextensible WLC model in the regimes of low ($f<15$ pN, green line) and high ($f>15$ pN, black line) forces. The fitting values are $p=1.04(8)$ nm and $l=0.612(7)$ nm, and $p=0.79(4)$ nm and $l=0.655(6)$ nm for the low- and high-force fits, respectively. The fitting parameters for the persistence and contour length ($p$, $l$) for each molecule are shown in Figure~\ref{fig:3}b. 

The obtained values for $p$ and $l$ for the LHs (yellow circles) are compatible and in agreement with those obtained for the SH at high forces (black triangles). In contrast, at low forces (green triangles), data for the SH show that $l$ decreases and $p$ increases, matching the results without BSO (magenta triangles). For comparison, we show the range of values for $p$ and $l$ reported in the literature (pink and yellow bands) for several molecules~\cite{camunas2016reviewnucleic}, which are in agreement with ours. The results for SH with BSO show that the differences in the {obtained} elastic parameters between short and long molecules are due to the {complementary} fitting force regimes rather than the different molecular lengths. We have also fitted data to the extensible WLC model (Equation~(\ref{eq::wlc_extensible})) that interpolates between the two elastic behaviours and reproduces the ssDNA elasticity throughout the whole range of forces (Figure~\ref{fig:3}a, blue dashed line), obtaining $l=0.618(14)$ nm, $p=0.95(4)$ nm, $Y=1100(200)$ pN. 

In general, an accurate knowledge of the ssDNA elastic properties is essential to extrapolate force-spectroscopy measurements at a finite force with zero-force bulk measurements. In particular, the free energy of stretching of ssDNA is crucial to derive the free energies of the different nearest-neighbors (NN) motifs from unzipping experiments~\cite{Huguet2010PNAS,Huguet2017NAR}. This free energy of stretching at force $f$ can be computed as the area below the measured FECs $f(x)$ (Figure~\ref{fig:3}c, inset) defined as $\Delta G^{\rm str}=\int^{x}_{x_0}{f(x)\,dx}$, where $x_0$ and $x$ are the initial and final ssDNA extensions. The estimation of the free energy of stretching, for the different fitted models, is shown in Figure~\ref{fig:3}c. As we can see from the figure, differences in $\Delta G^{\rm str}$ increase with force. At 15 pN (the typical force value of unzipping experiments), the relative error between the inextensible WLC fit (black and green lines) and the extensible WLC one (blue dashed line) is about $\sim $5\%. We also show the stretching contribution obtained {with} the FJC (continuous magenta line) and the extensible-FJC (dashed magenta line). At 15 pN, the highest error ($\sim$15\%) is committed for the two-parameter FJC {fit} ($c=0.570(4)$ nm and $b=1.36(5)$ nm). This was the model used for high salts in Reference~\cite{Huguet2010PNAS} to derive the NN basepair energy parameters from unzipping experiments. In that case, the committed error was about $0.1$ kcal/mol or $\sim$15\%. The extensible FJC, with three fitting parameters ($c=0.512(10)$ nm, $b=1.75(10)$ nm, $S=360(70)$ pN), overestimates the stretching energy at 15 pN by only 5\% with respect to the extensible WLC. 

\begin{figure}[h!]
\includegraphics[width=0.95\linewidth]{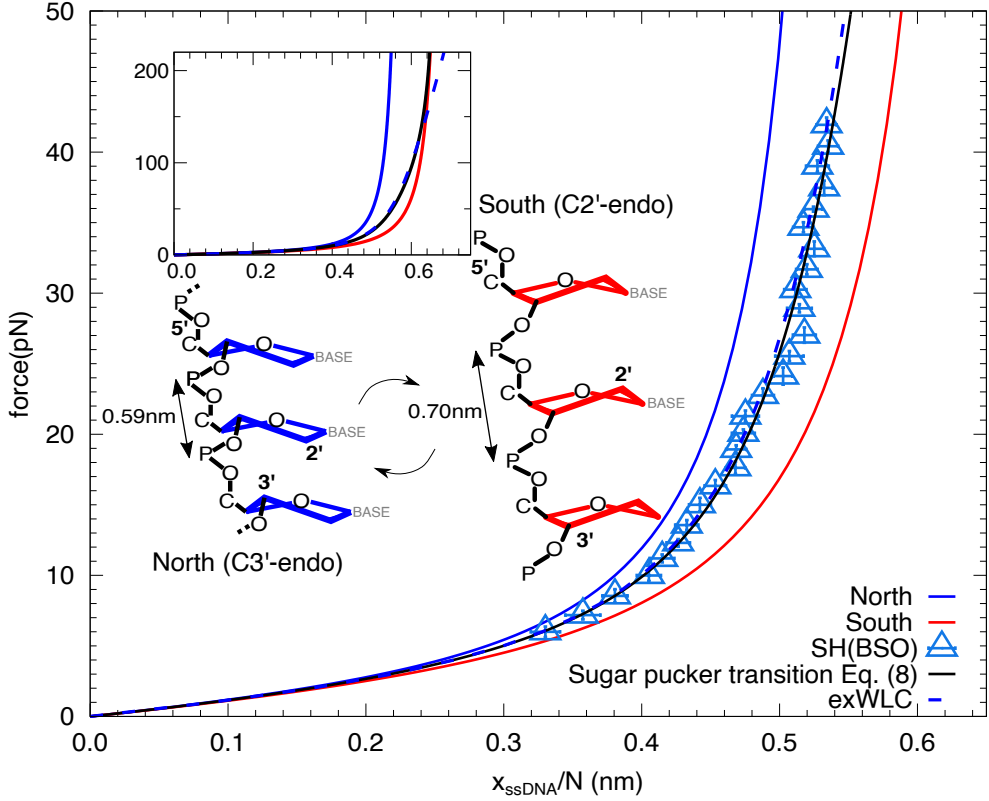}
\caption{\scriptsize{
Sugar pucker transition. Prediction by the NS model Equation (\ref{eq:FEC_model}) (black line) compared to the extensible WLC (dashed blue line) fit to the FEC of SH, obtained the BSO method (empty blue triangles, also shown in Figure \ref{fig:3}a). Blue and red lines correspond to the inextensible WLC FEC Equation (\ref{eq::wlc_marko_siggia}) obtained by imposing the crystallographic contour length of the south and north configurations (schematic depiction shown), respectively, and the persistence lengths obtained from fitting the data to the NS model ($p_{\rm S}=0.86(4)$ nm and $p_{\rm N}=0.98(4)$ nm, and a free energy of puckering per base of $\Delta G^{\rm pucker}=2J=0.38(4)$ kcal/mol). The error of the fits are obtained from bootstrapping. (Inset) Same as in the main figure, but in a higher force range, showing that forces as high as 150 pN (all bases are in the south configuration) must be reached to see significant deviations from the NS model and the extensible WLC.}
}
\label{fig:4}
\end{figure}

The fact that the low- and high-force regimes can be fitted to two inextensible WLC models --with different elastic parameters-- suggests the existence of a force-induced transition in the elasticity of ssDNA. This transition occurs between a ssDNA conformation of large persistence length and short contour length at low forces and a ssDNA conformation of shorter persistence length and longer contour length at high forces. The extensible WLC model reproduces the transition between the two behaviors. Interestingly, the contour length per base $l$ of the two different fitted force regime (0.612(14) nm, low forces; 0.655(6) nm, high forces) are close to the interphosphate distances of the two sugar pucker conformations of ssDNA (0.59 nm, C3$'$-endo; 0.70 nm, C2$'$-endo)~\cite{saengerprinciples2013,sinden1994dna}.
These are the north (C3$'$-endo) and south (C2$'$-endo) conformations, shown in Figure~\ref{fig:4}. The extensible WLC can be interpreted in terms of a non-cooperative transition between the two conformations. 
To model this transition we introduce a minimal North-South (NS) model for a chain of $N$ nucleotides under an externally applied force $f$ described by the following Hamiltonian:

\begin{equation}
\mathcal{H}(\sigma_i,f)=\left[J -\int_{0}^{f}\frac{\Delta x(f')}{2}df'\right]\sum_{i}\sigma_i-N\int_{0}^{f}\overline{x}(f')df',
    \label{eq:hamiltonian}
\end{equation}
where $\sigma_i$ represents the state of the nucleotide in the $i$th position ($1\leq i\leq N$), which can take $+1/-1$ values, corresponding to the south/north conformations. The elasticity of a single nucleotide in the south or north conformations are described by $x_S(f)$ and $x_N(f)$, respectively, and are modelled using the inextensible WLC (Equation~(\ref{eq::wlc_marko_siggia}) with different contour lengths $l_S$ and $l_N$. Equation~(\ref{eq:hamiltonian}) contains the average extension \mbox{$\overline{x}=(x_S(f)+x_N(f))/2$} and the difference in extension $\Delta x(f)=(x_S(f)-x_N(f))$ of the two conformations. The parameter $J$ is related to the free-energy difference per base between the south and north conformations as $\Delta G^{\rm pucker}=2J$. Note that for $J>0$ the north conformation is favored at zero force. 

The extension of the system described in Equation~(\ref{eq:hamiltonian}) is given by:

\begin{equation}
x_{\rm ssDNA}(f)=N\left[\overline{x}(f)+\frac{\Delta x(f)}{2}\tanh{\left(\frac{\int_{0}^{f}\frac{\Delta x(f')}{2}df'-J}{2k_BT}\right)} \right].
    \label{eq:FEC_model}
\end{equation}

We take the contour length of each conformation equal to the crystallographic one ($l_N=0.59$ nm and $l_S=0.70$ nm~\cite{sinden1994dna}) and fit Equation~(\ref{eq:FEC_model}) (black line, Figure~\ref{fig:4}) to the data for the SH in the whole range of forces (empty blue triangles, Figure~\ref{fig:4}). We obtain $J=0.19(2)$ kcal/mol, $p_N=0.98(4)$ nm and $p_S=0.86(4)$ nm. The coexistence force between the two conformations is about $\sim$40 pN. Such a large force is due to the small difference of extensions between the two conformations, which is only $\sim$1 \AA.
In the inset we show the FEC up to 220 pN, showing that the transition is very smooth and 90\% of the bases are in the south conformation only above $\sim$100 pN (Figure~\ref{fig:4}, inset). This is due to the lack of cooperativity. For large cooperativity, the transition between the two inextensible WLC would be observed as a plateau in the FEC around the coexistence force. Notice that the extensible WLC (dashed blue line, main Figure~\ref{fig:4} and inset) fits perfectly the prediction of the model below $f<150$ pN.

\section{Discussion}

We have studied the elasticity of ssDNA molecules over three orders of magnitude in length (60 bases to 14 kbases). The FECs obtained for the six LH sequences studied ($348\leq N\leq 6840$ bp) collapse into a single curve. This shows that, at high forces impeding secondary structure formation ($\sim$15 pN), long molecules with similar GC content ($\sim$50\% ) can be reproduced with the same set of elastic parameters. These are different from those needed to reproduce the FEC of the SH at low forces ($\lesssim$15 pN). Nevertheless, the FEC of the SH, obtained using the BSO method matches that of the LHs at high forces. This demonstrates the absence of molecular length effects in the elastic response of ssDNA.  Therefore the inextensible WLC model fails to reproduce the ssDNA elasticity in the studied force range (4--40 pN). By adding an extensibility term to the WLC, {we successfully} reproduce the experimental FECs throughout the force range. This result has implications in force-spectroscopy experiments for DNA, where a precise estimation of the ssDNA stretching energy is necessary to  derive the energetics of many molecular reactions (unzipping, unwinding, etc.). The committed error in the estimation of stretching free energies can be up to $\sim$15\%, when using the inextensible WLC or the FJC.


We have interpreted the observed extensibility in the FEC as a force-induced change in the conformation of the sugar pucker. The energy difference between the two sugar pucker conformations is $2J\sim0.4$ kcal/mol and falls in the range of values reported in the literature, {both} in computational~\cite{Wang2000,Arora2003} and semi-empirical~\cite{Nam2007,Gaus2013,Huang2014} studies. These values vary from  $\sim$0.5 to $\sim$$-$0.75 kcal/mol for different nucleotides, favoring either the north or south conformations, respectively. Interestingly, our measurements with {the} SH show that the north conformation is favored at zero force. This is also the natural conformation of dsRNA, where the sugar pucker is in the A-form~\cite{neidle2008}.  
This transition is different from the stacking-unstacking transition observed in nucleic acids. Recent pulling studies in homopolymeric ssDNA and ssRNA sequences~\cite{mcintosh2014,seol2007elasticity} have shown a stacking-unstacking transition where the inter-base distance nearly doubles between 0 and 25 pN. This transition is cooperative, showing a force plateau in the FECs around 15 pN. The magnitude of the contour length change and the cooperativity observed differ from the transition described in this work. In fact, sequences showing stacking contain a high fraction of consecutive purines, which is not the case for our hairpins, that are essentially random.

Evidence of the sugar pucker transition could be obtained by directly observing the two conformations in real time. For example, one might think of a hopping experiment in the passive mode where the force fluctuates on a sufficiently short ssDNA oligo ($\lesssim$10 bases) while monitoring the force. A resolution of about $\sim$1 \AA  ($\sim$10 fN) is needed to discriminate the different force levels, which is currently out of reach. Moreover, given the low energy difference between the two conformations, $2J\sim0.4$ kcal/mol, the transition would be very fast, probably beyond the corner frequency of the bead. The latter increases with the pulling force, therefore hopping experiments should be carried out at high forces to discriminate sugar pucker transitions in real time. This is a challenge for future experiments. Further support to our model might be obtained by pulling ssDNA tethers above $\sim$100 pN, to check whether data agrees with the FEC predicted by the model. At these very high forces, each sugar pucker conformation may also have some degree of extensibility. 
Future experiments may also address the case of ssRNA where the north conformation is more stable than for~ssDNA.

\section{Materials and Methods}
\label{S:Materials}
\subsection{Optical Tweezers Setup}
The experiments are carried out using a miniaturized dual-beam setup~\cite{Smith2003,Zaltron2020,Gieseler2021}. Briefly, two tightly focused counter-propagating laser beams (P = 200 mW, $\lambda$ = 845 nm) create a single optical trap, modelled as an harmonic potential (Figure~\ref{fig:1}a). Experiments are performed by tethering a DNA molecule  between two polystyrene beads, one is captured in the optical trap whereas the other is held by air suction in the tip of a glass micropippete.
Tethers are made by dsDNA tails (handles) that are labeled with one or several Biotins or Digoxigenins  which can bind selectively to Streptavidin (2.1 $\upmu$m Kisker Biotech) or anti-digoxigenin (3.0--3.4 $\upmu$m Kisker Biotech) coated beads, respectively. Experiments are performed in a microfluidics chamber where a micropippete is placed and used to immobilize the 2.1 $\upmu$m bead by air suction.~The force exerted on the optically trapped bead is determined by directly measuring the change of light momentum using Position Sensitive Detectors (PSD). The position of the optical trap is determined by diverting $\sim$5\% of each laser beam to a secondary PSD. The instrument has a resolution of 0.1 pN and 1 nm at a 1 kHz acquisition rate.

\subsection{DNA Substrates}
ssDNA is {generated} by mechanically unfolding DNA hairpins (see next section for details). Two types of hairpin are used, shown schematically in Figure~\ref{fig:1}b. Long hairpins (LH), denoted as HN, have a stem of N basepairs ($348 \leq {\rm N}\leq 6840$) and a tetraloop. The short hairpin (SH) corresponds  to a 20 bp-stem hairpin with a 20 bases loop. 
Two different preparations for the DNA substrates are done, described in detail in {the Supplementary Material}. SH is synthesized by annealing and ligating a different set of oligonucleotides, while LHs are synthesized using a long DNA fragment obtained either by a polymerase chain reaction amplification or by a digestion of the linearized $\lambda$-phage DNA. The hairpins are annealed to dsDNA handles that are used as spacers. Labeling of the handles is achieved by a digoxigenin/biotin tailing using a terminal transferase. All experiments are performed at 25 {\celsius} in a buffer containing 10 mM Tris pH 7.5, 0.01\%NaN$_3$, 10 mM MgCl$_2$.

\subsection{Measuring the ssDNA Molecular Extension}
\label{subsec:molecular_extension}
For the long hairpins (LHs) ssDNA is obtained using the blocking-loop oligonucleotide (BLO) method as described in~\cite{bosco2014elastic}. Briefly, a 25--30 bases oligonucleotide, complementary to the loop region, prevents the re-zipping of the molecule once it is fully unzipped (force-distance curves, FDC, represented Figure~\ref{fig:2}a). To measure the ssDNA elasticity we flow oligonucleotide-free buffer to the central channel to remove the excess of oligonucleotide. Several pulling and relaxing cycles are recorded, by moving the optical trap at constant speed ($\sim$50 nm/s). At a given force $f$, the measured trap position, $\lambda$,  is related to the molecular extension of the stretched ssDNA, $x_{\rm ssDNA}(f)$, by

\begin{equation}
x_{\rm ssDNA}(f)= \lambda(f)-x_h(f)-x_b(f)-x_{0},
\label{eq:lambda_long}
\end{equation}
where $x_h(f)$ is the extension of the handles  (characterized in~\cite{forns2012handles}), $x_b(f)$ is the displacement of the bead from the center of the optical trap, and $x_0$ is a shift of the trap position relative to the position detector (schematically shown in Figure~\ref{fig:1}a, right). A typical force-extension curve (FEC) for a long ssDNA molecule, $f(x)$, is shown in Figure~\ref{fig:2}b. The experimental FEC for each LH is obtained by averaging 4--10 molecules, each one containing 4--10~cycles.

For the SH, the BLO method alters significantly the ssDNA elasticity and an alternative approach is required. We use the two branches methods that is based on the generation of a large hysteresis between the stretching and relaxing {FDC}s by using a long loop (20 bases long). The ssDNA extension $x_{\rm ssDNA}$ can be measured directly from the differences in the trap position between the folded and unfolded force-{distance} branches at a given force $f$, $\lambda_U-\lambda_F$, as:

\begin{equation}
x_{\rm ssDNA}(f)=\lambda_{U}(f)-\lambda_{F}(f)+x_{d}(f)+\lambda_0(t),
\label{eq:ssDNA_short}
\end{equation}
where $x_d(f)$ is the diameter of the hairpin projected along the stretching direction (see next section) and $\lambda_0(t)$ is the correction of the drift using a spline interpolation after the alignment of the cycles (described in~\cite{Viader2021}). In Figure~\ref{fig:2}d, we show the FEC for the SH using this method, obtained by averaging 6 molecules, each one with 30--100 cycles.  The range of forces in which the FEC can be measured is limited by the unfolding ($\sim$15 pN) and the refolding ($\sim$4 pN for a 20 b-loop) forces.

In order to measure the FEC for SH at larger forces, the experiments are performed with a blocking-splint oligonucleotide (BSO) that is fully complementary to one flanking ssDNA handle (29b) and partially (15 bases) to the other DNA handle (as schematically depicted in Figure~\ref{fig:2}e). A spacer of 4 Thymines ($\sim$2 nm in length) is inserted between the 29 b and 15 b segments to properly hybridize the oligos to the flanking ssDNA handles. The BSO acts as an staple and stabilizes the folded state of the hairpin up to high forces $f<f_r$. The particular value of $f_r$ depends on the total length of the BSO. In our case, a BSO with an overhang of 15 bases gives $f_r\sim 40$pN, high enough to extract $x_{\rm ssDNA}(f)$ over a wide range of forces. An analysis similar to that shown in Equation~(\ref{eq:ssDNA_short}) gives:

\begin{equation}
x_{\rm ssDNA}(f)+x_{h2}^{15,\rm ssDNA}(f)=\lambda_{U}(f)-\lambda_{F}(f)+x_{d}(f)+x_{h2}^{15,\rm dsDNA}(f)+\lambda_0(t),
\label{eq:ssDNA_short_block}
\end{equation}
where $x_{h2}^{15,\rm dsDNA}(f)$ and $x_{h2}^{15,\rm ssDNA}(f)$ correspond respectively to the extension of $15$ bp dsDNA and 
15 bases ssDNA at force $f$. Notice that the l.h.s. of Equation~(\ref{eq:ssDNA_short_block}) corresponds to the combined ssDNA extension of the 60 bases of the unfolded SH and the 15 ssDNA bases of handle 2. In Figure~\ref{fig:2}f we show the FEC for the SH using the BSO method obtained by averaging 7 molecules, each one with 5--30 cycles. {Throughout this work, the force versus the trap position will be referred as force-distance curve (FDC), while the force versus molecular extension is designed as force-extension curve (FEC). The latter is obtained from the former via Equations~(\ref{eq:lambda_long})--(\ref{eq:ssDNA_short_block}).}

\subsection{DNA Elastic Models}
\label{Subsec:elastic_models}

The elastic response of bio-polymers has been mainly described with two models: the Freely-Jointed Chain (FJC)  and the the Worm-Like Chain (WLC)~\cite{marko1995stretching,Smith1996, bouchiat1999estimating}.  The former describes the polymer as
$N$ rigid bonds of length $c$ that can orient freely in the space.
In the FJC model, the end-to-end extension of the polymer $x$ as a function  the applied force $f$ reads as: 

\begin{equation}
x(f)=N\,c\left( \coth\left( \frac{b f}{k_BT} \right)-\frac{k_BT}{b f}\right).
\label{eq:FJC}
\end{equation}

Where $b$ is the Kuhn length, $T$ is the temperature and $k_B$ is Boltzmann's constant. The FJC has been used to model the hairpin orientation (introduced in Section~\ref{subsec:molecular_extension}), with  $c=b=2$ nm, corresponding to the double-helix diameter~\cite{saengerprinciples2013, sinden1994dna}. An extension of this model, the so-called extensible Freely-Jointed Chain, includes 
bond-stretching, by modifying the extension $x$ of Equation~(\ref{eq:FJC}) as $x_e(f)=x(f)\left(1+f/S\right)$, where $S$ is the stretch~modulus.

The WLC describes the polymer as a flexible rod. Whereas the FJC model of \mbox{Equation~(\ref{eq:FJC})} is purely entropic, the WLC adds an enthalpic contribution which takes into account a bending penalty. The WLC has two characteristic lengths: $l$, the contour length per base and $p$, the persistence length. The latter is defined as the distance along which the tangent vector de-correlates. 
The model is not analytically solvable, but an interpolation formula for the low- and high-force regimes is given by~\cite{marko1995stretching}:

\begin{equation}
f=\frac{k_BT}{p}\left(\frac{1}{4}\left(1-\frac{x}{n\,l}\right)^{-2}-\frac{1}{4}+\frac{x}{n\,l}\right),
\label{eq::wlc_marko_siggia}
\end{equation}
where  $n$ is the total number of bases, the total contour length of polymer being $L=n \times l$. 
Both models have been used to fit the FEC of ssDNA. For persistence lengths that are comparable to the interphosphate distance ($p\sim c$) the two models are equivalent with \mbox{$b=2p$}. The results reported in the literature are compatible with this relation (\mbox{$0.5\leq p\leq 2$~nm}, $1\leq b\leq 4$ nm)~\cite{adamcik2006observation,Smith1996,camunas2016reviewnucleic}.
The WLC can also be extended to include a contour length that increases with the force. This formulation leads to the so called extensible WLC model, which is described by the following force-extension relation~\cite{Wang1997}:
\begin{equation}
f=\frac{k_BT}{p}\left(\frac{1}{4}\left(1-\frac{x}{n\,l}+\frac{f}{Y}\right)^{-2}-\frac{1}{4}+\frac{x}{n\,l}+\frac{f}{Y}\right),
\label{eq::wlc_extensible}
\end{equation}
where $Y$ stands for the Young modulus of the chain and the other parameters have been defined above.

\appendix
\section{DNA hairpin synthesis}
\label{S:Ap.DNAsynth}

The short hairpin (SH) was synthesized following the protocol in Ref.~\cite{alemany2014determination} (named as CD4L20). The oligonucleotide forming the 3' end of the hairpin was labelled with a digoxigenin tailing. After a purification step (\textit{QIA Nucleotide removal kit}), all the oligonucleotides forming the hairpin (Table \ref{Ap.table:oligos_short}) were annealed by starting at a high temperature (\SI{70}{\celsius}) and \SI{1}{\celsius} was decreased every minute until room temperature was reached. The hairpin was next ligated using the T4 DNA ligase (New England Biolabs) in an overnight reaction (\SI{16}{\celsius}). 

\begin{figure}[h!]
\centering
\includegraphics[width=\linewidth]{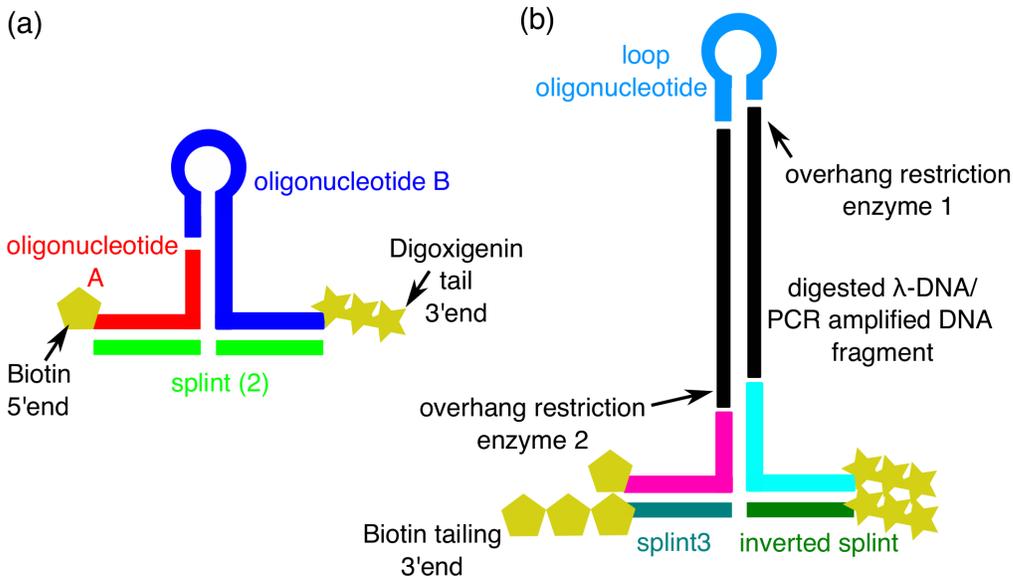}
\caption{{\bf Hairpin synthesis.} (a) The SH is assembled by ligating two oligonucleotides (blue and red). The oligonucleotide A is purchased biotinylated and the oligonucletide B is end-labelled with digoxigenins using the T4 terminal transferase. The splint oligonucletides are annealed to create dsDNA handles.
(b) The LHs are assembled by ligating a set of oligonucleotides (magenta, cyan, blue) to the PCR-amplified and digested $\lambda$-phage fragment (black). Note that the complementary strands of the handles are also tailed. Color code as in Tables~\ref{Ap.table:oligos_short} and \ref{Ap.table:oligos_long}.  }
\label{fig:S:hairpin}
\end{figure}

The long hairpins (LHs) were synthesized following a procedure based either on a PCR amplification of a dsDNA segment or digesting a segment of the linearized $\lambda$-DNA.
$\textrm{H}700$, $\textrm{H}964$, $\textrm{H}4452$ and $\textrm{H}13680$ were prepared as described in Refs.~\cite{camunas2015footprinting} and \cite{camunas2013electrostatic}.
Finally, $\textrm{H}1904$ and $\textrm{H}7138$ were synthesized following the protocol in Ref.~\cite{camunas2013electrostatic}, changing the restriction enzyme for the digestion step: {\sl EcoRI} (New England Biolabs) for $\textrm{H}7138$, and {\sl BspHI} (New England Biolabs) for $\textrm{H}1904$. The sequences of the oligonucleotides used for preparing the DNA hairpins are given in Sec.~\ref{S:Ap.DNAoligos}. Fig.~\ref{fig:S:hairpin}(a)-(b) shows an scheme of the molecular construct for the SH and LH hairpins, respectively.

\clearpage
\newpage

\section{Oligonucleotides for hairpin synthesis}
\label{S:Ap.DNAoligos}
\begin{table}[h!]\footnotesize
\centering
\begin{tabularx}{\textwidth}{s b}
\hline
\textbf{Name} & \textbf{Sequence} \\
\hline
\textcolor{red}{SH-A} & \textcolor{red}{5$'$-Biotin-AGT TAG TGG TGG AAA CAC AGT GCC AGC GCG AAC CCA CAA ACC GTG ATG GCT GTC CTT GGA GTC ATA CGC AA -3$'$}  \\
\textcolor{blue}{SH-B} & \textcolor{blue}{5$'$-GAA GGA TG\textbf{G AAA AAA AAA AAA AAA AAA A}CA TCC TTC TTG CGT ATG ACT CCA AGG ACA GCC ATC ACG GTT TGT GGG TTC AGT TAG TGG TGG AAA CAC AGT GCC AGC GC-3$'$} \\
\textcolor{green}{splint} & \textcolor{green}{5$'$-GCG CTG GCA CTG TGT TTC CAC CAC TAA CT-3$'$}\\
\hline

\end{tabularx}
\caption{\label{Ap.table:oligos_short}Oligonucleotides used for the synthesis of the SH. The loop region is shown in bold.}
\end{table}

\begin{table}[h!]\footnotesize
\centering
\begin{tabularx}{\textwidth}{s b}
\hline
\textbf{Name} & \textbf{Sequence} \\
\hline
\textcolor{myblue}{13680b-loop} & \textcolor{myblue}{5$'$-Pho-GAT CGC CAG TTC GCG TTC GCC AGC ATC CG{\bf A CTA} CGG ATG CTG GCG AAC GCG AAC TGG C-3$'$} \\
\textcolor{myblue}{7138b-loop} & \textcolor{myblue}{5$'$-Pho-AAT TGC CAG TTC GCG TTC GCC AGC ATC CG{\bf A CTA} CGG ATG CTG GCG AAC GCG AAC TGG C-3$'$} \\
\textcolor{myblue}{4452b-loop} & \textcolor{myblue}{5$'$-Pho-TGA TAG CCT {\bf ACT A}AG GCT ATC ACA TG-3$'$} \\
\textcolor{myblue}{1904b-loop} & \textcolor{myblue}{5$'$-Pho-CAT GAC AGT CGT TAG TAA CTA ACA TGA TAG TTA C{\bf TT TT}G TAA CTA TCA TGT TAG TTA CTA ACG ACT GT-3$'$}  \\
\textcolor{myblue}{964b-loop} & \textcolor{myblue}{5$'$-Pho-GTC ACT TAG TAA CTA ACA TGA TAG TTA C{\bf TT TT}G TAA CTA TCA TGT TAG TTA CTA A-3$'$} \\
\textcolor{myblue}{700b-loop} & \textcolor{myblue}{5$'$-Pho-GTC ACT TAG TAA CTA ACA TGA TAG TTA C{\bf TT TT}G TAA CTA TCA TGT TAG TTA CTA A-3$'$} \\
\textcolor{mypink}{Bio-cosRshort} & \textcolor{mypink}{5$'$-Bio-GAC TTC ACT AAT ACG ACT CAC TAT AGG GAA ATA GAG ACA CAT ATA TAA TAG ATC TT-3$'$} \\
\textcolor{mycyan}{cosRlong} & \textcolor{mycyan}{5$'$-Pho-GGG CGG CGA CCT AAG ATC TAT TAT ATA TGT GTC TCT ATT AGT TAG TGG TGG AAA CAC AGT GCC AGC GC-3$'$} \\
\textcolor{mypink}{Bio-cosLshort}  & \textcolor{mypink}{5$'$-Bio-GAC TTC ACT AAT ACG ACT CAC TAT AGG GAA ATA GAG ACA CAT ATA TAA TAG ATC TT-3$'$} \\
\textcolor{mycyan}{cosLlong}  & \textcolor{mycyan}{5$'$-Pho-AGG TCG CCG CCC AAG ATC TAT TAT ATA TGA GTC TCT ATT AGT TAG TGG TGG AAA CAC AGT GCC AGC GC 3$'$}\\
\textcolor{mypink}{HandBio-SMFP} &\textcolor{mypink}{ 5$'$-Bio-GAC TTC ACT AAT ACG ACT CAC TAT AGG GAA ATA GAG ACA CAT ATA TAA TAG ATC TTC GCA CTG AC -3$'$}\\
\textcolor{mycyan}{HandDig-SMFP} & \textcolor{mycyan}{5$'$-Pho-AAG ATC TAT TAT ATA TGT GTC TCT ATT AGT TAG TGG TGG AAA CAC AGT GCC AGC GC -3$'$}\\
\textcolor{mydarkgreen}{splint3}  & \textcolor{mydarkgreen}{5$'$-TCC CTA TAG TGA GTC GTA TTA GTG AAG TC-3$'$} \\
\textcolor{mygreen}{inverted-splint} &  \textcolor{mygreen}{3$'$-AAA AA-5$'$-5$'$-GCG CTG GCA CTG TGT TTC CAC CAC TAA C(SpC3)-3$'$}\\
\hline
\end{tabularx}
\caption{\label{Ap.table:oligos_long}Oligonucleotides used for the synthesis of long DNA hairpins. The loop region is shown in bold.}
\end{table}

\begin{table}[h!]
\centering
\begin{tabularx}{\textwidth}{s b}
\hline
\textbf{Name} & \textbf{Sequence} \\
\hline
13680b-block-loop & 5$'$-TAG TCG GAT GCT GGC GAA CGC GAA CTG GCG-3$'$  \\
7138b -block-loop & 5$'$-TAG TCG GAT GCT GGC GAA CGC GAA CTG GCG-3$'$\\
4452b-block-loop & 5$'$-TAG TAG GCT ATC ACA TGC TGG CCA CCG GCT-3$'$ \\
1904b-block-loop & 5$'$-TTA CAA AAG TAA CTA TCA TGT TAG T-3$'$ \\
964b-block-loop & 5$'$-TTA CAA AAG TAA CTA TCA TGT TAG T-3$'$ \\
700b-block-loop & 5$'$-TTA CAA AAG TAA CTA TCA TGT TAG T-3$'$ \\
\hline
\end{tabularx}
\caption{\label{S:Ap.table:oligos_block}Blocking loop oligonucleotides used to generate ssDNA FECs for the different LHs.}
\end{table}

\newpage
\clearpage

\bibliographystyle{unsrt}

\end{document}